\documentclass[aps, prx, twocolumn,superscriptaddress,longbibliography,numerical]{revtex4-1}

\usepackage{amsmath}
\usepackage{epsfig}
\usepackage{graphicx}
\usepackage{bm}
\usepackage{amssymb}
\usepackage{slashed}
\usepackage{hyperref}
\hypersetup{
     colorlinks   = true,
     citecolor    = blue,
     urlcolor = blue,
     linkcolor = blue
}

\usepackage{float}
\usepackage{subcaption}

\newcommand{\bsigma}{{\boldsymbol \sigma}}
\newcommand{\bnabla}{{\boldsymbol \nabla}}

\newcommand{\e}{\text{e}}
\newcommand{\bDelta}{{\boldsymbol \Delta}}
\newcommand{\br}{\mathbf{r}}
\newcommand{\U}{\text{U}}

\begin{document}

\title{Antichiral and nematicity-wave superconductivity}

 \author{Mats Barkman}
 \affiliation{Department of Physics, Royal Institute of Technology, SE-106 91 Stockholm, Sweden}

 \author{Alexander~A.~Zyuzin}
  \affiliation{Department of Applied Physics, Aalto University, P.~O.~Box 15100, FI-00076 AALTO, Finland}
\affiliation{Ioffe Physical--Technical Institute,~194021 St.~Petersburg, Russia}

 \author{Egor Babaev}
 \affiliation{Department of Physics, Royal Institute of Technology, SE-106 91 Stockholm, Sweden}


\begin{abstract}
 Larkin-Ovchinnikov superconducting state has spontaneous modulation of Cooper pair density, while Fulde-Ferrell state has a
 spontaneous modulation in the phase of the order parameter.
 We report that 
 a quasi-two-dimensional Dirac metal, under certain conditions has principally different inhomogeneous superconducting  states that by contrast have spontaneous modulation in a submanifold of a multiple-symmetries-breaking order parameter.
 The first state we find can be viewed as a nematic superconductor where the nematicity vector spontaneously breaks rotational and translational symmetries due to spatial modulation. The other demonstrated state is a chiral superconductor with spontaneously broken time-reversal and translational symmetries. It is characterized by an order parameter, which forms a lattice pattern of alternating chiralities.
\end{abstract}
\maketitle
For most superconductors, the ground state represents a configuration where the superconducting fields
are homogeneous and can be classified according to the pairing symmetries.
A generalization was  theoretically proposed by 
  Larkin and Ovchinnikov \cite{Larkin_Ovchinnikov} and independently by Fulde and Ferrell \cite{Fulde_Ferrell}.
  It was demonstrated that not only $U(1)$ symmetry can be broken in such a superconducting state but
also the translational symmetry due to formation of Cooper pairs occurring with finite momentum. That state is called Larkin-Ovchinnikov-Fulde-Ferrell (LOFF) state.
In the simplest case, it can be caused by the pair-breaking effect of the Zeeman field in conventional superconductors.
There are other mechanisms for the formation of inhomogeneous states in different systems such as cold atoms \cite{Kinnunen2018}, or dense quark matter in neutron star interiors \cite{alford2001crystalline}. This made periodically modulated superconducting and superfluid states a subject of wide interest (for reviews see  Refs. \cite{Buzdin_Review, Shimahara_Review}).

In this Rapid Communication we discuss 
a class of materials that supports inhomogeneous states which are
principally different from  the LOFF solutions.
Namely, we find an inhomogeneous counterpart of the chiral superconducting state, where the system spontaneously forms a pattern of alternating chiralities, thereby breaking 
both translational and time reversal symmetries. Since the  time reversal 
shares $\mathbb{Z}_2$ symmetry with Ising magnets,    we term this state as ``antichiral" state in analogy to the antiferromagnetism.
Another state we find is an inhomogeneous counterpart of the nematic superconducting state where the nematic vector is modulated, forming a nematicity-wave.

We show that these states occur 
for the type of microscopic physics like that found in the recently discovered doped topological insulators \cite{Exp_BiSe1, Exp_BiSe2, Exp_BiSe3}. Experimental studies of these materials suggest the presence of nematicity in the superconducting state with two components and odd-parity symmetric order parameter. However the chiral state might be more energetically favorable in the quasi-two-dimensional limit of these Dirac materials, in which the Fermi surface is cylindrical. The type of the superconducting pairing and the Majorana surface states are subjects of intense investigation and debate \cite{Fu_Berg, Fu_nematic, Debate1, Debate2}.  For a review of the superconducting instabilities in these materials, see Ref. \cite{TopSC_review}.

We consider superconductivity in a Dirac metal. We begin by demonstrating that  an inhomogeneous state can be realized there in the absence of external field, by the violation of inversion symmetry, different from the Zeeman pair-breaking mechanism.  Next we present a microscopic derivation
of a multicomponent Ginzburg-Landau (GL) model for a superconductor with imbalanced fermionic populations. Then
we numerically find solutions that minimize the microscopically derived free-energy GL functional.  By that we find two kinds of inhomogeneous superconducting states in which the chirality or the nematicity of the order parameter is spatially modulated.

Let us start with a microscopic model of the quasi-two-dimensional Dirac metal with the cylindrical Fermi surface. 
This model might, for example, be applied  to the doped $\textrm{Bi}_2\textrm{Se}_3$  topological insulator with a layered crystal structure, in which the Fermi surface might become quasi-two-dimensional under doping \cite{Cylinder_FS}. The quintuple layer unit cell in this material is modeled by a bilayer structure in which the helical electronic states occupying the top and bottom layers are hybridized. The low-energy excitations in the system can be described by the Hamiltonian $\mathcal{H}_0=\int_\mathcal{S} \text{d}^2 r \Psi^{\dag}(\mathbf{r})H_0(\mathbf{r})\Psi(\mathbf{r})$, where $\mathcal{S}$ is the area of the system and
\begin{equation}\label{Main}
H_0(\mathbf{r})=
-iv \left[\sigma_x \partial_y - \sigma_y \partial_x \right]\tau_z -V\tau_z+m\tau_x ,
\end{equation}
in which $v$ is the Fermi velocity characterizing the two-dimensional Dirac dispersion, $m$ is the spin-conserving tunnel matrix element between two orbitals describing the mass of the Dirac fermion, $\sigma_j$ and $\tau_j$ with  $j=x,y,z$ are the Pauli matrices describing the spin and the orbital degrees of freedom respectively, and
$
\Psi(\mathbf{r})=(\Psi_{\uparrow,1}(\mathbf{r}),\Psi_{\downarrow,1}(\mathbf{r}),\Psi_{\uparrow,2}(\mathbf{r}),\Psi_{\downarrow,2}(\mathbf{r}))^{\mathrm{T}}$ is the electron operator ($\uparrow, \downarrow$ and $1, 2$ define two spin projections and two orbitals within the unit cell, respectively). Finally,  we  will be using $\hbar =1$ units and suppress the explicit spin and orbital indices for clarity of notation.
\begin{figure}[t]
\center
\begin{subfigure}[t]{0.23\textwidth}
		\center
        \includegraphics[width=\textwidth]{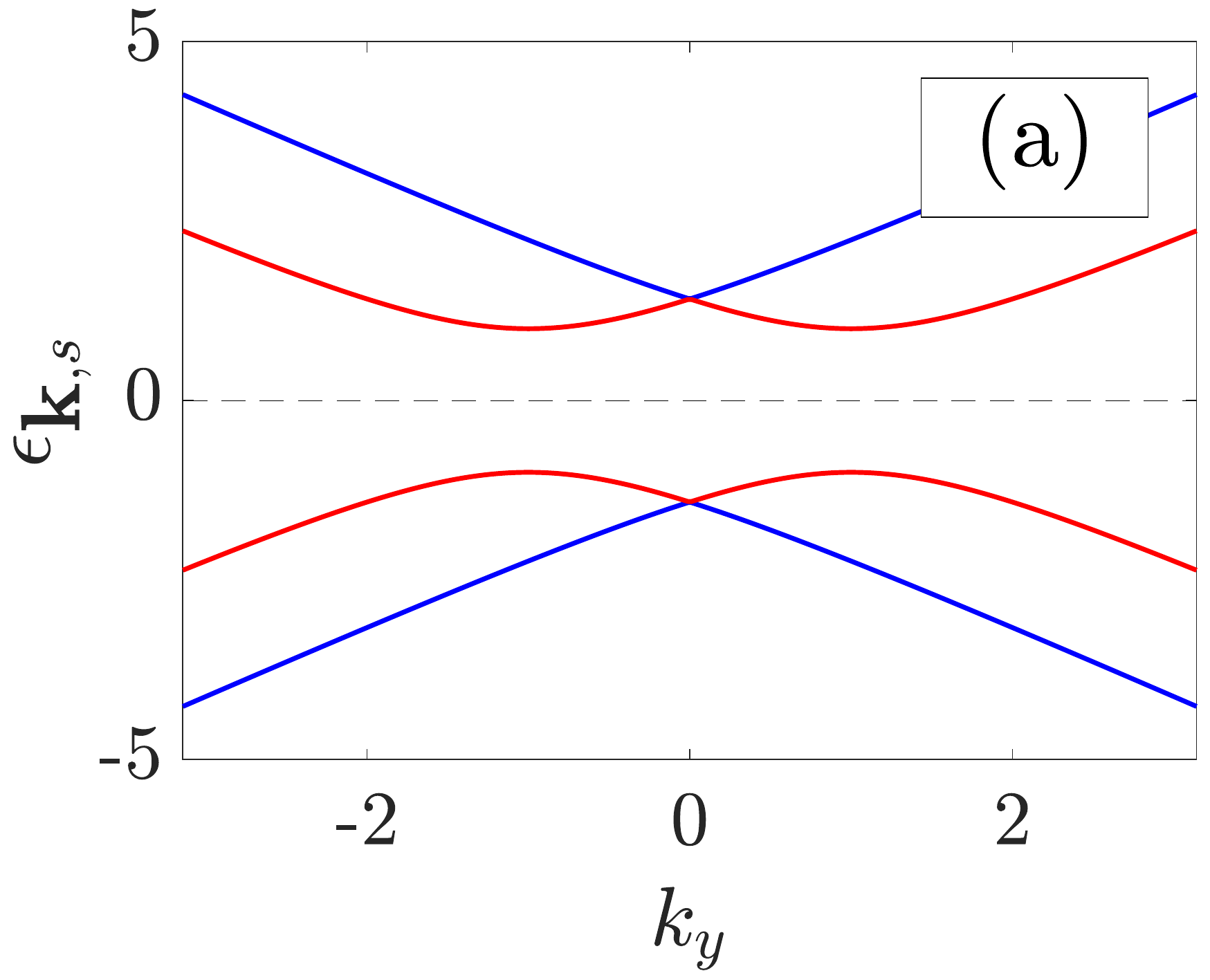}
\end{subfigure}
\begin{subfigure}[t]{0.23\textwidth}
		\center
        \includegraphics[width=\textwidth]{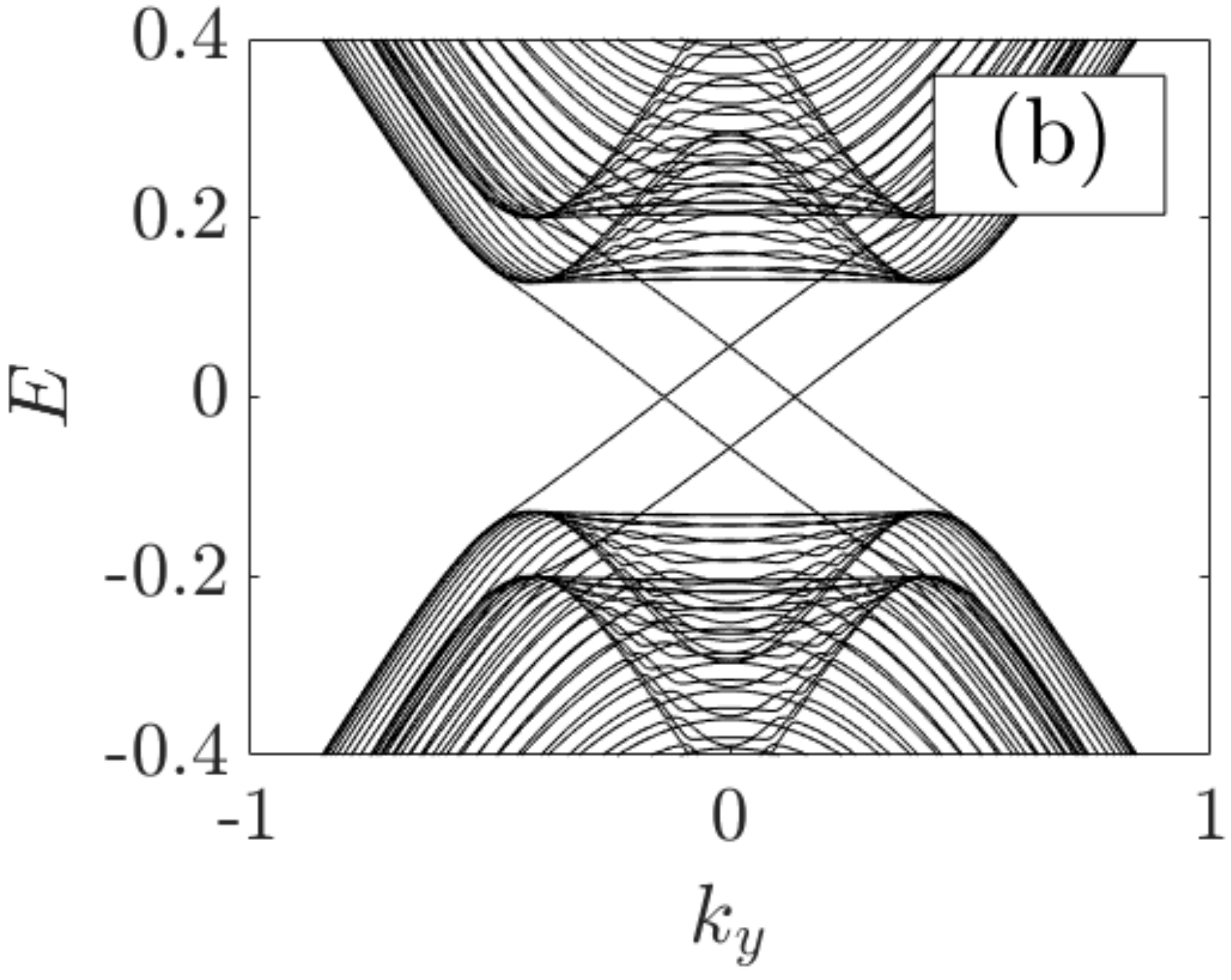}
\end{subfigure}
\caption{(a) The spectrum of particles $\epsilon_{\mathbf{k},s }$ with $k_x=0$ in the normal state of the system for a given mass $m$ in the case when $V\neq 0$. (b) The eigenstate dispersion for chiral phase of the quasi-two-dimensional superconductor along the $y$ direction in momentum space for a sample of finite width along the $x$ direction in the situation, in which the 
inversion-symmetry breaking parameter is $V=0$. The bulk spectrum is gapped and each surface hosts a pair of unidirectional chiral modes. The bulk gap closes at $V \approx |\Delta_c|$.} \label{fig1}
\end{figure}

The term $V\tau_z$ 
 violates the inversion symmetry, and corresponds to an electrostatic potential difference between the orbitals, or to the external bias, provided the system is in the two-dimensional bilayer limit (for more discussion of this model see, for example, Ref. \cite{Franz}). Although there are many terms which break inversion, we restrict ourselves only to inversion-breaking term independent of momentum, which preserves rotational symmetry around the $z$ axis. 
We then assume that the Fermi level is in the conduction band and consider the Fermi energy $\mu$ to be the largest energy scale in the system, i.e., we set $\mu > \sqrt{m^2 +V^2}$. 

At $V=0$, the bulk of the metal is inversion symmetric due to the presence of two orbitals with opposite signs of the Fermi velocity within the unit cell. 
At finite $V$, the spectrum of quasiparticles (without the dispersion along the $z$ axis) is given by 
$
\epsilon_{\mathbf{k},s } = \sqrt{m^2+(vk+  s V)^2},
$
where $s=\pm$ and $k = (k_x^2+k_y^2)^{1/2}$. The role of the $V\tau_z$ term is to lift the double degeneracy at every momentum except at the inversion symmetric $k_x=k_y=0$ point, see  Fig. \ref{fig1}.
At $m > \mu - m > V$ the spectrum is parabolic $\epsilon_{\mathbf{k},s} = m+ v^2k^2/2m +s V\sqrt{1-m^2/\mu^2}$, whereas at $\mu-m > V>m$ the spectrum is linear $\epsilon_{\mathbf{k},s} = vk + s V + m^2/2\mu$. 
We argue that the phase volume of the inhomogeneous state is larger in the latter case, which we will adopt in what follows.

Consider now the superconductivity in the system described above.
Among possible superconducting instabilities \cite{TopSC_review}, we focus on the singlet-interorbital and spin-triplet pairing. 
The BCS mean-field Hamiltonian, in Nambu representation, measured from the chemical potential, is given by
$\mathcal{H}=\frac{1}{2}\int_\mathcal{S} \text{d}^2 r\Phi^{\dag}(\mathbf{r})H(\mathbf{r})\Phi(\mathbf{r})$,
where $\Phi^{\dag}(\mathbf{r})=(\Psi^{\dag}(\mathbf{r}),
\Psi^{\textrm{T}}(\mathbf{r}))$ is the Nambu operator in the superconducting state.
The Hamiltonian density of the system is given by
\begin{equation}\label{BCS_HAM}
H(\mathbf{r}) = \bigg[
 \begin{matrix}
  H_0(\mathbf{r})-\mu & i \sigma_y  \tau_y  (\bsigma \cdot \bDelta )\\
  -i \sigma_y  \tau_y  (\bsigma \cdot \bDelta^* ) & - H_0^{*}(\mathbf{r})+\mu
 \end{matrix}
\bigg],
\end{equation}
where $\bsigma = (\sigma_x, \sigma_y)$ and with the vector $\mathbf{\Delta} = ( \Delta_x, \Delta_y )$ composed of two components of the order parameter,
in which $\Delta_x \propto i (\langle \Psi_{\uparrow,2}\Psi_{\uparrow,1} \rangle - \langle \Psi_{\downarrow,2} \Psi_{\downarrow,1} \rangle)$ and $\Delta_y \propto \langle \Psi_{\uparrow,2}\Psi_{\uparrow,1} \rangle + \langle  \Psi_{\downarrow,2} \Psi_{\downarrow,1} \rangle$. For the complete classification of the homogeneous order parameters in this kind of systems we refer the reader to Ref. \cite{Fu_Berg} and to a review article \cite{TopSC_review}. In the absence of inversion breaking, it is the uniform chiral state with the two-fold degenerate order parameter $\mathbf{\Delta}_{\pm} = \Delta_c (1, \pm i)$ which is energetically more favorable than the nematic state $\mathbf{\Delta}\propto (\cos \theta, \sin \theta)$ with constant nematic angle $\theta$, as was noticed, for example, in Ref. \cite{Zyuzin_Skyrmion} and then later extended in Ref. \cite{Chirolli_chiral}.

The uniform chiral state spontaneously breaks time-reversal symmetry.
The close analogies of this chiral state are the quasi-two-dimensional $p$-wave superconductors and the A phase in superfluid $\mathrm{^3He}$ \cite{Anderson_Morel, Balian_Werthamer}, where the spin-triplet pairing might be stabilized by the spin-fluctuation feedback mechanism \cite{Anderson_Brinkman}.

We ``project"  the multiband Hamiltonian in Eq. \ref{Main} onto a $2 \times 2$ subspace corresponding to the conduction band and arrive at the effective BCS Hamiltonian in momentum representation provided spatially homogeneous order parameter
\begin{equation}\label{BCS_diag}
H(\mathbf{k}) = \left[
 \begin{matrix}
  \xi_{\mathbf{k}} - \sigma_z V & i \sigma_y \sigma_z(\Delta_x \hat{k}_x + \Delta_y \hat{k}_y) \\
  i\sigma_y \sigma_z(\Delta^*_x \hat{k}_x + \Delta^*_y \hat{k}_y)& -\xi_{\mathbf{k}} +\sigma_z V
 \end{matrix}
\right],
\end{equation}
where $ \xi_{\mathbf{k}}  =vk - \mu$ (in which we have already included corrections $\propto m^2/\mu$ into the chemical potential) and $\hat{\mathbf{k}}$ is the unit vector in the direction of momentum. 
The spectrum of bulk quasiparticles is given by
$
E_{s,\pm}(\mathbf{k}) = s V \pm \sqrt{\xi_{\mathbf{k}}^2 + |\Delta_x \hat{k}_x + \Delta_y \hat{k}_y|^2 }, s=\pm
$.
It is seen that at $V=0$, the superconductor is gapped. The boundary of the superconductor hosts pairs of one-way propagating Andreev-Majorana modes (see Fig. \ref{fig1}, and for review , see \cite{Silaev_Volovik}).
The increase of $V$ closes the gap in the spectrum. The system becomes gapless provided $V \geq |\Delta_c| $,
where the superconducting state can become unstable toward the transition to the spatially inhomogeneous phase.

We will now investigate the two-component chiral superconductor in the presence of inversion breaking within the microscopically derived GL formalism.
The role of the Zeeman pair-breaking effect is played by the spatial inversion, which removes the orbital degeneracy, leading to a mismatch between the Fermi surfaces and hence destroys the interorbital superconducting coupling.

The standard LOFF state microscopic derivation of the Ginzburg-Landau functional was presented in e.g. \cite{Buzdin_Kachkachi}. The key, state-defining feature 
of  a modulated state is negative sign of second-order gradients, which is responsible for the formation
of an inhomogeneous state. Because of the negative second-order gradient terms, it is necessary to retain
fourth-order derivative terms in GL expansions for LOFF states. 
Furthermore, in some of the microscopic models the prefactors of the  fourth-order potential terms
also become negative, which in turn requires retaining potential terms up to sixth order.  We apply the standard procedure to derive the GL functional from the microscopic model under consideration. To this end we integrate out the fermionic degrees of freedom using the BCS Hamiltonian in Eq. \ref{BCS_HAM} and utilizing the normal-state Green's function $G(\omega_n,\mathbf{k})=[i\omega_n+\mu-H_0(\mathbf{k})]^{-1}$, where $\omega_n=\pi T(2n+1)$ is the fermionic Matsubara frequency, $T$ is the temperature, and $n \in \mathbb{Z}$.
Keeping also the six-order terms, we derive GL functional $\mathcal{F} = \int_{\mathcal{S}}\text{d}^2 r [F_2+F_4+F_6] $ where for readability the free-energy density has been separated into three groups, classified according to the powers of the gap field. The second-order terms in the free-energy density are given by 
\begin{align}
F_2= &   
 \sum_{s} \bigg[\alpha |\Delta_s|^2+ \beta |\bnabla \Delta_s|^2+ \delta |\bnabla^2 \Delta_s|^2\bigg] + \\\nonumber
& \frac{\beta}{2} \bigg[| \partial_x \Delta_x + \partial_y \Delta_y |^2 - |\partial_x \Delta_y - \partial_y \Delta_x|^2\bigg] + \label{Chiral_GL} \\
& \frac{2\delta}{3}\left[ \left|\bnabla(\partial_x \Delta_x + \partial_y \Delta_y )\right|^2 - \left|\bnabla(\partial_x \Delta_y - \partial_y \Delta_x)\right|^2 \right], \nonumber
\end{align}
where the fourth-order gradient terms are included to ensure that the free energy is bounded from below when second-order gradient terms become negative. 
Explicitly, $|\mathbf{\Delta}|^2 = |\Delta_x|^2+|\Delta_y|^2$ and $ \mathbf{\Delta} \times \mathbf{\Delta}^* = \Delta_x\Delta_y^*- \Delta_y\Delta_x^*$. 
We retain the following terms at the fourth order in fields  
\begin{eqnarray}\label{Chiral_GL4}\nonumber
&F_4&=    \gamma |\mathbf{\Delta}|^4- \frac{\gamma}{3}  |\mathbf{\Delta} \times \mathbf{\Delta}^*|^2 
+  \eta |\boldsymbol{\Delta}|^2[ |\bnabla \Delta_x|^2 + |\bnabla \Delta_y|^2]
\\\nonumber 
&+& \frac{\eta}{3}|\boldsymbol{\Delta}|^2 \left[| \partial_x \Delta_x + \partial_y \Delta_y |^2 - |\partial_x \Delta_y - \partial_y \Delta_x|^2\right]
\\\nonumber
&-&
\frac{\eta}{3}\bigg\{
|\Delta_x\boldsymbol{\nabla}\Delta_y^* - \Delta_y\boldsymbol{\nabla}\Delta_x^*  |^2 + |\Delta_x\boldsymbol{\nabla}\Delta_y - \Delta_y\boldsymbol{\nabla}\Delta_x |^2\\\nonumber
&-&
(|\Delta_x|^2-|\Delta_y|^2 )(|\partial_x\boldsymbol{\Delta}|^2- |\partial_y\boldsymbol{\Delta}|^2 )- (\Delta_x\Delta_y^*+\Delta_x^*\Delta_y) \\
&\times& (\partial_x\Delta_x^*\partial_y\Delta_x +\partial_x\Delta_y^*\partial_y\Delta_y +\mathrm{c.c.} )
\bigg\}.
\end{eqnarray}
In addition we need to include potential
terms at sixth order to ensure that the free energy
is always bounded from below:
\begin{eqnarray}\label{Chiral_GL6}
&F_6&=   \varepsilon \bigg[ |\mathbf{\Delta}|^4- \frac{3}{5} |\mathbf{\Delta} \times \mathbf{\Delta}^*|^2 \bigg] |\mathbf{\Delta}|^2.
\end{eqnarray}
The coefficients in the GL functional density are given by $ \alpha = - \pi \nu_2[K_1(V)-K_1(V_0) ]$, 
$
\beta = \frac{\pi \nu_2 v^2}{8}K_3(V), 
\gamma = \frac{3 \pi \nu_2}{16} K_3(V), 
\delta = -\frac{3}{8}\frac{\pi \nu_2 v^4}{16} K_5(V),
\eta =  - \frac{3 \pi \nu_2 v^2}{16} K_5(V),
\varepsilon = -\frac{5\pi \nu_2}{64} K_5(V)$, 
in which $\nu_2 = \mu /8\pi v^2$ is the two-dimensional density of states at the Fermi level per spin and orbital in the massless limit and $
K_{j}(V) =  2 T \mathrm{Re}\sum_{n=0}^{\infty}\left(\omega_n - i V\right)^{-j},~~j\geq 1$, where $V_0$ is the parameter corresponding to the transition into uniform superconducting state.

For the case $V=0$, where there is no imbalance between the orbitals in Eq. \ref{Main}, the coefficients $\beta$ and $\gamma$ in the GL functional are positive and hence the terms with coefficients $\delta, \eta, \epsilon$ can be neglected. The GL functional for this case was analyzed, for example, in Ref. \cite{Venderbos_Fu}.

Importantly, the sign in front of the quartic term $|\mathbf{\Delta} \times \mathbf{\Delta}^*|^2 = 4|\Delta_x|^2 |\Delta_y|^2 \sin^2 \phi$, where $\phi$ is the phase difference between $\Delta_x$ and $\Delta_y$, in Eq. \ref{Chiral_GL4} chooses whether the superconductor is in the chiral phase with spontaneous broken time-reversal symmetry and $\phi = \pm \pi/2$ or in the nematic phase with broken rotational symmetry and $\phi =0$. 

The negative sign here results in the two degenerate chiral phases, where GL functional density is minimized by the uniform order parameters $\boldsymbol{\Delta}_{\pm} = \Delta_c(1, \pm i)$, with $|\Delta_c|=\sqrt{-3\alpha/8\gamma}$. 
 
It is important to note, however, that there is a phase transition between the chiral and nematic phases as a function of the anisotropy of the Fermi surface \cite{Zyuzin_Skyrmion, Chirolli_chiral}. The evolution of the Fermi surface in the microscopic model Eq. \ref{Main} from the cylindrical to elliptic leads to a sign flip in front of the term $|\mathbf{\Delta} \times \mathbf{\Delta}^*|^2$. Indeed in a three-dimensional isotropic model the GL functional density is minimized by the nematic order parameter $\boldsymbol{\Delta} \propto (\cos \theta, \sin \theta)$ up to arbitrary nematic angle $\theta$ \cite{Venderbos_Fu}.

Now we will see that increasing $V$ leads to decreases of $\beta$ and $\gamma$
and the situation changes dramatically. Coefficients $\beta$ and $\gamma$ simultaneously change sign signaling the instability of the uniform superconductivity toward the formation of an inhomogeneous state. To ensure that the free energy is bounded from below, the terms with positive coefficients $\delta$, $\eta$, and $\varepsilon$ should be retained in the GL functional in the regime where coefficients $\beta$ and $\gamma$ are negative. 

The resulting states are investigated in two dimensions numerically by minimizing the free energy functional $\mathcal{F}$, adopting periodic boundary conditions, using the finite element method framework FreeFem++ \cite{FreeFem} and the nonlinear conjugate gradient method. 
In the regime where $\beta$ and $\gamma$ are negative, it is convenient to rescale the model  by defining the dimensionless quantities $\tilde{\Delta}_s = \Delta_s/|\bDelta_\U|$, $\tilde{\alpha}=\alpha/\alpha_\U $, $\tilde{\br}=q_0 \br$ where $|\bDelta_\U |= -\gamma/2\varepsilon$, $\alpha_\U = \gamma^2/4\varepsilon$ and $q_0^2 = -\beta/2\delta$. The free energy $\mathcal{F}=  \alpha_\U |\bDelta_\U|^2/q_0^2 \cdot \tilde{\mathcal{F}}$ where the rescaled free energy $\tilde{\mathcal{F}}$ is identical to the original, having replaced $\Delta_s$ with $\tilde{\Delta}_s$, $\alpha$ with $\tilde{\alpha}$ and so on, where $\tilde{\gamma}=-2\tilde{\varepsilon}=-2$, $\tilde{\beta}=-2\tilde{\delta} = -2 (\beta/\gamma)^2 \cdot \varepsilon/\delta$ and $\tilde{\eta} = \beta/\gamma \cdot \eta/\delta$. All coefficients are constant except $\tilde{\alpha}$, which parametrizes both $V$ and $T$.
The full order parameter can be parametrized as 
\begin{equation}
\begin{aligned}
\boldsymbol{\Delta}(\br) = & \left(|\Delta_x (\br)| \e^{i \phi_x (\br)},|\Delta_y (\br)| \e^{i \phi_y (\br)}\right) = 
 \\
 & |\boldsymbol{\Delta}(\br)| \e^{i \chi(\br)} \left( \e^{i \phi (\br)}\cos \theta(\br) ,  \sin \theta (\br) \right), 
\end{aligned}
\end{equation}
where $|\Delta_s (\br)|$ and $\phi_s (\br)$ are the amplitudes and phases of each component.  We introduce  $|\boldsymbol{\Delta}(\br)|$ and $\chi(\br)$ as the amplitude and the phase of the overall  order parameter prefactor, respectively, $\phi (\br)$ as the phase difference  between its components and $\theta(\br)$ as the nematic angle.

\begin{figure}[t]
\center
\includegraphics[width=0.49\textwidth]{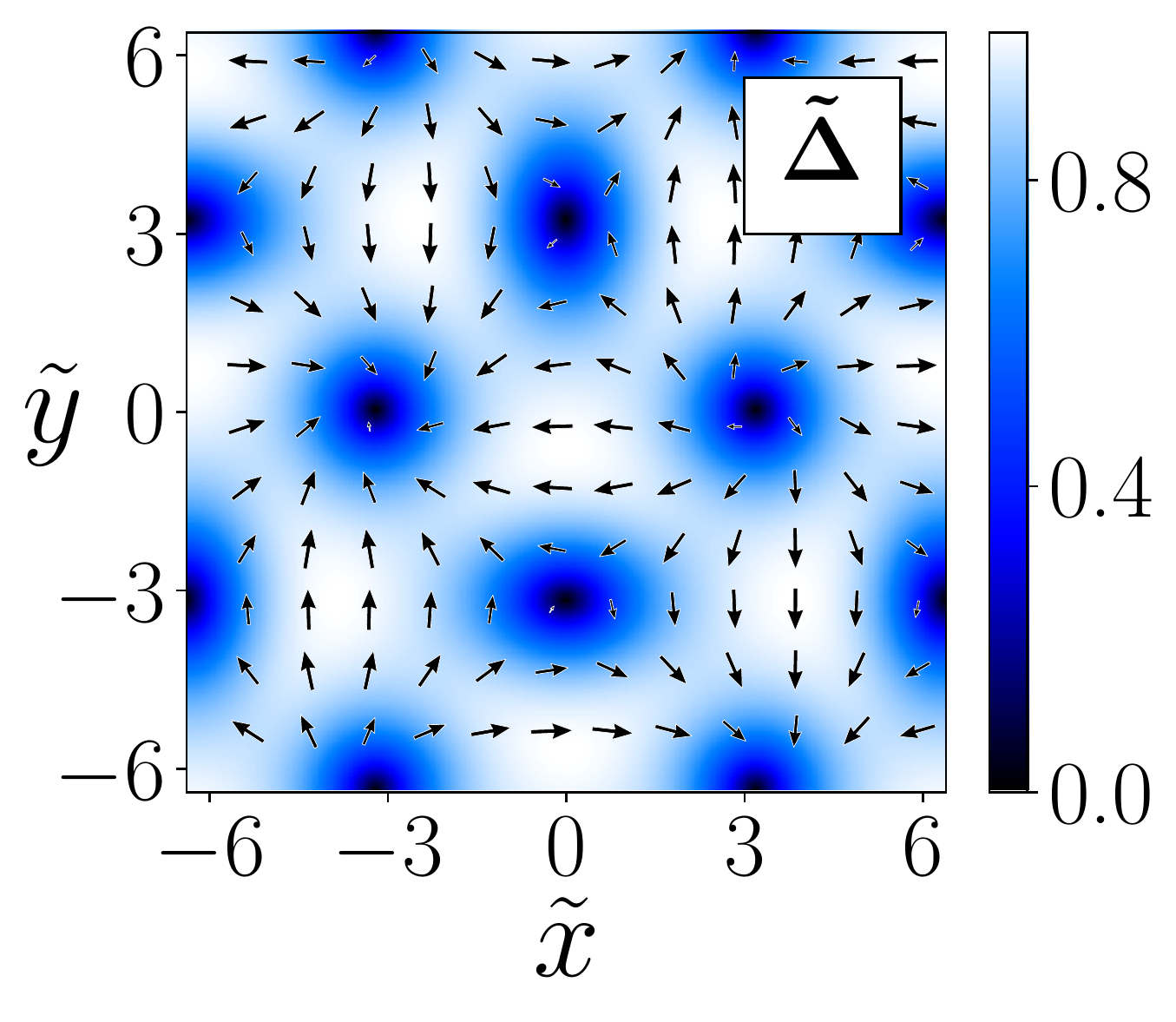}
\caption{Inhomogeneous state obtained by numerical minimization of the rescaled free-energy functional $\tilde{\mathcal{F}}$. The superconducting order parameter $\tilde{\bDelta}=(\tilde{\Delta}_x,\tilde{\Delta}_y)$ is shown as the vector field and the modulus $|\tilde{\bDelta}|$ as the heat map. The state exhibits modulation in relative density, and a periodic modulation of the nematic angle $\theta (\br)$, forming a nematic vortex-antivortex lattice.  We term this state ``nematicity-wave". The parameter $\tilde{\alpha}$ which parametrizes both temperature and inversion breaking is set to $\tilde{\alpha}=1$, at which the nematicity-wave state is energetically preferable over the uniform state.} \label{fig: Anti-nematic}
\end{figure}

We find a stable state which is characterized by modulation in relative density between the two components $\Delta_x$ and $\Delta_y$. The state represents a nematicity wave as shown in Fig. \ref{fig: Anti-nematic}. 
The structure of the nematic angle $\theta (\br)$, indicated by the corresponding nematic vector field, shows the formation of a nematic vortex-antivortex lattice.
 Correspondingly the uniform superconducting state becomes unstable toward the transition into such a  new type of inhomogeneous state. 

\begin{figure}[t]
\center
\includegraphics[width=0.49\textwidth]{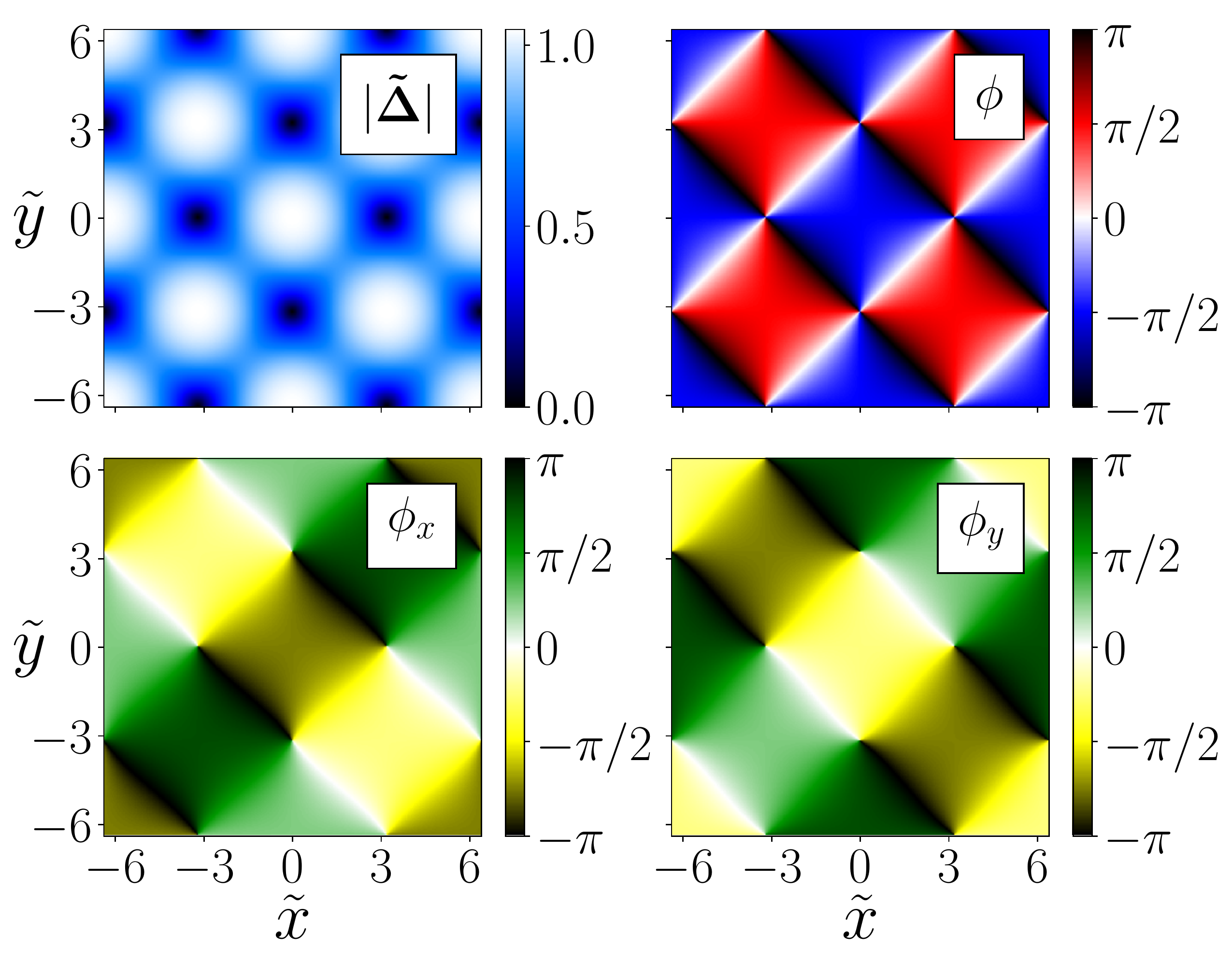}
\caption{Inhomogeneous state obtained by numerical minimization of the free-energy functional, where the term proportional to $|\bDelta \times \bDelta ^*|^2$ has been replaced by $\left\lbrace \gamma + \frac{3 \varepsilon}{2} |\bDelta|^2 \right\rbrace |\bDelta \times \bDelta ^*|^2$ (i.e. where the potential terms would produce
the nematic superconducting state if the gradient terms were positive) and the parameter $\tilde{\alpha}=1$. The phase difference in $\phi$ shows that the state has the form of the pattern with  alternating chiralities. Like the uniform chiral state, this inhomogeneous state breaks $\mathbb{Z}_2$ symmetry associated with flipping chiralities in each segment. We call this state ``antichiral".
The individual phases $\phi_x$ and $\phi_y$ show that the pattern 
leads to the formation of a lattice of   cocentered vortices in one component and antivortices in the other component.} \label{fig: Anti-chiral}
\end{figure}

Next we investigate a similar GL functional where the prefactor to $|\bDelta \times \bDelta ^*|^2$ has been modified. 
Although this change of prefactor is phenomenological, it can be used to qualitatively describe the nematic state: in fact we have shown that it is indeed favorable over the chiral in three dimensions, with inhomogeneous phase arising from the different depairing term $V \tau_y \sigma_z $. 
This term is replaced by $\left\lbrace \gamma + \frac{3 \varepsilon}{2} |\bDelta|^2 \right\rbrace |\bDelta \times \bDelta ^*|^2$, such that in the regime where $\beta$ and $\gamma$ are positive, the energy is minimized by the uniform gap parameter $\bDelta_\theta = \Delta_n ( \cos \theta, \sin \theta)$ where $|\Delta_n| = \sqrt{-\alpha /2\gamma}$ and the nematic angle $\theta$ is constant. When the gradient terms are positive such a uniform state
is called nematic (see, e.g.  \cite{Venderbos_Fu, Zyuzin_Skyrmion}). By contrast, in the regime where $\beta$ and $\gamma$ are negative we find a state with inhomogeneous gap parameter which is energetically preferred over the uniform state. It is characterized by modulation in the phase difference $\phi$ and a formation of a checkerboard pattern of alternating chirality.  The state is shown in Fig. \ref{fig: Anti-chiral}. 

 Studying the individual phases $\phi_x$ and $\phi_y$ one can see that the formation of this pattern is accompanied by the vortex-antivortex lattice.
 Namely, in the corners  of each chirality domain there forms a vortex in one component and an antivortex in the other component.
Since these vortices and antivortices are co-centered this composite vortex does not carry magnetic flux.

It is worthwhile to note that the homogeneous nematic and chiral phases are topological, in the sense that they support Majorana Kramer's pairs and chiral Majorana modes bound at the defects (e.g., at the vortices and boundaries), respectively (see \cite{TopSC_review, Venderbos_Fu}). The properties of the topological modes in the inhomogeneous nematic and chiral states will be studied separately.

To summarize, we 
demonstrated two classes of inhomogeneous superconducting states:
the first ``nematicity-wave" is a superconducting state
where there is a periodically modulated nematic order parameter.
The second class is ``antichiral" where the system forms an alternating pattern of opposite chiralities. In an analogy with 
an Ising antiferromagnet it has broken time-reversal $\mathbb{Z}_2$ symmetry associated with the flipping of chiralities, or equivalently
shifting the pattern by half of its period. 
 
The   patterns that we find are accompanied by a spontaneous formation of vortex-antivortex lattice.
This  highly unconventional effect is energetically penalized for uniform superconductors with 
positive quadratic gradient terms, but becomes the very efficient energy minimizing  solution when
the quadratic gradient terms are negative.
We demonstrated microscopically, that the candidate materials where this state may form are doped topological insulators. 
It was  recently discussed that the phase transitions from modulated
to normal superconducting states
should proceed in multiple steps with a system first loosing
superconductivity in the bulk but only at elevated temperatures
on the surface \cite{BarkmanBabaev,SurfaceSamoilenka}.
It can be  utilized to distinguish these states  from the  homogeneous  chiral and nematic superconductors,
for example in calorimetry measurements. Since both the antichiral and nematicity-wave states exhibit lattice of density zeros due to the presence of vortex-antivortex lattice, they should be observable in tunneling microscopy experiments. \\ 

\textit{Note added in proof:} 
Recently, there appeared an interesting study that considers a different generalization of the LOFF state for systems to the $p_x + i p_y$ case \cite{Fradkin}. In contrast to our work, there the system has nonalternating chirality but forms inhomogeneous states of stripes of the alternating overall sign of the order parameter with a given chirality.

\begin{acknowledgements}
A.Z. was supported by the Academy of Finland. M.B. and E.B. were supported by the Swedish Research
Council Grants No. 642-2013-7837 and No.  VR2016-06122 and Goran  Gustafsson  Foundation  for  Research  in  Natural  Sciences  and  Medicine. Part of this work was performed at the Aspen Center for Physics, which is supported by the U.S. National Science Foundation Grant No. PHY-1607611.
 The computations were performed on resources provided by the
Swedish National Infrastructure for Computing (SNIC) at the National Supercomputer Center
 in Linkoping, Sweden.
\end{acknowledgements}

\bibliography{main.bib}

\end{document}